\documentclass[nofootinbib,preprintnumbers,amsmath,amssymb]{revtex4}
\usepackage{graphicx}
\usepackage{dcolumn}
\usepackage{bm}
\usepackage{amsfonts}
\usepackage{amsmath}
\usepackage{amssymb}
\usepackage{amsthm}
\usepackage{citesort}

\begin{document}


\title{Upper Bounds for the Number of Hamiltonian Cycles}

\author{Jinshan Zhang}
\email{zjs02@mails.tsinghua.edu.cn}

\affiliation{
Department of Mathematical Sciences, Tsinghua University, Beijing, China, 100084 
}%


\begin{abstract}
\textbf{Abstract.} An upper bound for the number of Hamiltonian
cycles of symmetric diagraphs is established first in this paper,
which is tighter than the famous Minc's bound and the
Br$\acute{e}$gman's bound. A transformation on graphs is proposed,
so that counting the number of Hamiltonian cycles of an undirected
graph can be done  by counting the number of Hamiltonian cycles of
its corresponding symmetric directed graph. In this way, an upper
bound for the number of Hamiltonian cycles of undirected graphs is
also obtained.

\


\end{abstract}


\maketitle

\section{Introduction}
Let $G=(V,E)$ be an undirected graph with a vertex set
$V=\{v_{1},v_{2},\cdots,v_{n}\}$ and edge set $E$. An edge from
$v_{i}$ to $v_{j}$ is denoted by  $(v_{i},v_{j})$. For simplicity,
the vertices are  also denoted as $V=\{1,2,\cdots, n\}$.  A
Hamiltonian cycle of $G$ is a closed path that visits each of the
vertex once and only once. Similarly, if $G$ is a directed graph, a
closed directed path which visits each of the vertex once and only
once is a Hamiltonian cycle of a directed graph. In this paper, we
use the notation $v_{1}v_{2}\cdots v_{n}v_{1}$ and
$(v_{1},v_{2},\cdots,v_{n},v_{1})$ to denote Hamiltonian cycles in
undirected  and directed graphs respectively.

It is well known that the decision problem whether a graph contains
a Hamiltonian is NP-complete. Hence, counting the  number of
Hamiltonian cycles of a graph is a hard problem too.
G.A.Dirac\cite{DIR52} shows the existence of Hamiltonian cycles in
the undirected graph $G$ of minimum degree at least $(1/2
+\epsilon)n$, where $n$ is the number of vertices in $G$ and
$\epsilon > 0$. Counting the number of Hamiltonian cycles in such
graphs is still $\#$P-complete\cite{DFJ98}. Hence algorithms and
analysis are developed for approximating or estimating the number of
Hamiltonian cycles in both directed and undirected graphs . The best
asymptotic result of the number of Hamiltonian cycles on random
graphs is obtained by Janson\cite{JAN94}. N.Alon et. al. show better
lower and upper bounds of the maximum number of Hamiltonian cycles
in an $n$-tournament problem \cite{AAR01, ALON90}.
This paper mainly focuses on bounding the number of Hamiltonian
cycles. An upper bound of the number of Hamiltonian of arbitrary
symmetric directed graph is proposed. We also prove that the number
of Hamiltonian cycles of an undirected graph equals half of  that of
its corresponding symmetric direct graph. Hence, any bound on a
directed graph can be directly applied to bound the number of
Hamiltonian cycles of an undirected graph.

Our novel bound on a symmetric directed graph is better than one of
its natural bounds Minc's bound and tighter than Br\'{e}gman's bound
in many cases, e.g. when out-degrees are bounded by a constant
$K\leq 5$. Our proof of the new bound is mainly based on a random
algorithm for counting the number of Hamiltonian cycles on a
directed graph, which is modified from Rassmussen's
algorithm\cite{RAS94}. To apply the result on a symmetric directed
graph to the undirected graph, a very simple but useful
transformation that transforms counting the number of Hamiltonian
cycles of a undirected graph to that of a symmetric directed graph
is proposed.

The structure of this paper is as follows. Some nature bounds for
the number of Hamiltonian cycles led by matrix permanent are
introduced in section II. The Rassmussen's algorithm for counting
the number of Hamiltonian cycles is discussed and a modified
algorithm is presented in section III. Some fundamental properties
of the algorithms are given.  A new bound on a symmetric directed
graph is presented in section IV. A transformation extending the
result in symmetric direct graphs to undirected graphs is
established in section V. In this way, upper bounds of the number of
Hamiltonian cycles in undirected graphs is also obtained. Some
concluding remarks are proposed in section VI.

\section{Nature Bounds via Matrix Permanent}

To establish the results on the number of Hamiltonian cycles, some
related  concepts and results are introduced. Consider $G=(V,E)$ be
a directed graph with vertex set $V=\{v_{1},v_{2},\cdots,v_{n}\}$ and edge set $E$.
In the following section, our notations are only related to the directed graph
except in section V. \\\\
\textbf{Definition 2.1} Directed graph $G$ is called a symmetric
directed graph iff edge $(v_{i},v_{j})\in E$ $\Rightarrow$
$(v_{j},v_{i})\in E$, for any $i\neq
j$.\\\\
\textbf{Definition 2.2} An 1-factor of a directed graph  $G$ is a
spanning subgraph of $G$ in which all in-degrees and out-degrees are
1. \\

An example of the 1-factor is a spanning union of vertex disjoint
directed cycles. Let $NH(G)$ and $F(G)$ denote the number of
Hamiltonian cycles and the number of 1-factors of a graph $G$
respectively. Since every Hamiltonian cycle is also an 1-factor,
therefore
\begin{equation}
NH(G)\leq F(G)
\end{equation}
The permanent of a matrix $A = (a_{ij})_{n\times n}$ is defined as
\begin{equation}
\operatorname{Per}(A)=\sum\limits_{\sigma}\prod\limits_{i=1}^{n}a_{i\sigma(i)}
\end{equation}
where $\sigma$ goes over all the permutations $\{1,2,\cdots,n\}$.
The adjacency matrix $A = A_{G}$ of a graph $G$ is an $n$ by $n$ 0-1
matrix. The matrix $A = (a_{ij})$ is defined as $a_{ij} = 1$, if
$(v_{i}, v_{j}) \in E$; $a_{ij} = 0$, otherwise. Note the diagonal
entries of $A$ are all zero, and for any permutation $\sigma$,
$a_{i\sigma(i)}= 1$, $i=1,\cdots,n$ iff their corresponding edges in
$G$ form an 1-factor of $G$. Hence,
\begin{equation}
F(G)=\operatorname{Per}(A).
\end{equation}
Hence any upper bounds on matrix permanent would provide upper
bounds for the number of 1-factors and therefore the number of
Hamiltonian cycles.
For the permanent of a matrix, the following are two  famous upper bounds.\\\\
We present a similar definition as permanent of a matrix, and we
called it Hamilton of a matrix, which is defined as
$\operatorname{ham}(A)=a_{11}$, when $n=1$, and when $n\geq
2$,
\begin{displaymath}
\operatorname{ham}(A)=\sum\limits_{\{k_2,k_3,\cdots,k_n\}}a_{k_1k_2}a_{k_2k_3}
\cdots a_{k_{n-1}k_n}a_{k_nk_1},
\end{displaymath}
where $\{k_2,k_3,\cdots,k_n\}$ is over all the permutations of
$\{1,2,\cdots,n\}/\{k_1\}$ and $k_1$ is any number from the set
$\{1,2,\cdots,n\}$.\\

Considering the relation between $A$ and graph $G$, the elements in
the  set $\{a_{k_1k_2}, a_{k_2k_3}, \cdots a_{k_{n-1}k_n},
a_{k_nk_1}\}$ are all positive iff their corresponding edges in $G$
form a Hamiltonian cycle. Hence, $\operatorname{ham}(A)=NH(G)$.\\\\
\textbf{Theorem 2.1} (Minc's Bound) Let $A=(a_{i,j})$ be an $n
\times n$ 0-1 matrix with the row sum $r_{i}$, $i=1,2,\cdots,n$.
Then
\begin{equation}
\operatorname{Per}(A)\leq\prod\limits_{i=1}^{n}\frac{r_{i}+1}{2} .
\end{equation}
\textbf{Theorem 2.2} (Br\'{e}gman's Bound) Let $A=(a_{i,j})_{n
\times n}$ be an $n \times n$ 0-1 matrix, and $r_{i}$ denote the
number of ones in the row $i$, $1\leq i\leq n$. Then
\begin{equation}
\operatorname{Per}(A)\leq
\prod\limits_{i=1}^{n}(r_{i}!)^{\frac{1}{r_{i}}}.
\end{equation}

Due to Stirling formula $n!\leq\sqrt{2\pi n}n^n/e^{n-1/12}$, the
bound in Theorem 2.2 is tighter than that in Theorem 2.1. The bound
in Theorem 2.2 is conjectured by Minc in 1963 \cite{MIN63} and later
proved by Br\'{e}gman\cite{BREG73}. It plays an essential role in
the proof of the conjecture of Szele by N. Alon\cite{ALON90}. From
the formula (1)(3)(4)(5), we can naturally obtain an upper bound of
the number of Hamiltonian cycles of a directed graph. Particularly
it is an upper bound of the number of Hamiltonian cycles of a
symmetric directed graph.

\section{Modified Rassmussen's Algorithm}
In this section, We suppose $G=(V,E)$ be a directed graph with $n$
vertices $\{1,2,\cdots,n\}$ and $A=A_{G}=(a_{ij})_{n \times n}$ be
the adjacent matrix of the graph $G$. Let $A(ij)$ be the
$(n-1)\times (n-1)$ matrix obtained by removing row $i$ and column
$j$ from the matrix $A$; $A(i,:)$ is  row $i$ of the matrix $A$. For
any set $S$, let $|S|$ be the number of its elements. We now present
the algorithm given by Rassmussen in \cite{RAS94}.\\\\
\textbf{Algorithm 3.1}\\
\textbf{inputs}: $A$, an $n\times n$ 0-1 matrix;\\
\textbf{outputs}: $X_A$, the estimator of the number of Hamiltonian cycles in $G$;\\
\textbf{step0}: Let $p_i=0$, $i=1,\cdots,n$;\\
\textbf{step1}: For $i=1$ to $n$

If $|A(1, :)| = 1$; Set $p_n=a_{11}$, goto step2;

Else $W=\{j>1:a_{1j}=1\}$;

\ \ \ \ \ \ If $W=\emptyset$; Set $X_A=0$;

\ \ \ \ \ \ \ \ \ Stop;

\ \ \ \ \ \   Else

\ \ \ \ \ \ \ \ \ Choose $J$ from $W$ uniformly at random;

\ \ \ \ \ \ \ \ \ Let $p_{i} =|W|$;

\ \ \ \ \ \ \ \ \ Permutate the column $1$ and $J$;

\ \ \ \ \ \ \ \ \ Let $A = A(11)$;\\
\textbf{step2}: $X_A = p_1 \times\cdots\times p_n$.\\

Algorithm 3.1 presents an unbiased estimator of the number of
Hamiltonian cycles $G$,  which means the expectation of the output
$X_a$ is the number of Hamiltonian cycles in $G$. We present another
point of view of Algorithm 3.1. Through one random experiment of
Algorithm 3.1, if the output $Y_A$ is not zero, one obtains a
Hamiltonian cycle of $G$ by putting together all the edges
corresponding to the selected elements in A.
 Hence, each Hamiltonian cycle of $G$ can be selected with certain
probability. Suppose the Hamiltonian cycle $(1, k_1, \cdots,
k_{n-1}, 1)$ has been chosen, where $\{k_{i},i=1,2,\cdots,n-1\}$ is
a permutation of $\{2,3,\cdots,n\}$. In $i$th iteration of step1 in
Algorithm 3.1, the probability of some element $a_{ij}$ is selected
with probability
$$\frac{1}{p_{i}}, \ \ i=1,\cdots,n. $$
Hence this corresponding  Hamiltonian cycle formed by the edges
corresponding to the chosen elements is selected with the
probability $\frac{1}{p_1} \times\cdots\times \frac{1}{p_n}$ and the
output is $p_1 \times\cdots\times p_n$. From this viewpoint, the
output is an
unbiased estimator of the number of Hamiltonian cycles of $G$. We state it below.\\\\
\textbf{Theorem 3.1} Let $X_A$ the output of Algorithm 3.1 .
Then $E(X_A)=NH(G)$.\\\\
\textbf{Proof:} Let $\mathcal{H}(i)$ be one selected Hamiltonian
cycle and $Y_{\mathcal {H}(i)}$ denote the output when $\mathcal
{H}(i)$ is selected. From the above analysis, we see $\mathcal
{H}(i)$ can be selected with probability $\frac{1}{Y_{\mathcal
{H}(i)}}$. Hence,
\begin{displaymath}
E(X_A)=\sum\limits_{i=1}^{NH(G)}Y_{\mathcal
{H}(i)}\frac{1}{Y_{\mathcal {H}(i)}}=NH(G)
\end{displaymath}gives the result. $\Box$\\

Note that in one random experiment in Algorithm 3.1 in step1, the
element is selected by the ascending order of the row, or
equivalently say, we select the first element from the first row of
A, then select the second element from the second row of A and so
on. If we select the element by another fixed order of the row, we
can obtain Algorithm 3.2, which performs an essential part in
getting new upper bound in a symmetric directed graph.  In the
following Algorithm 3.2, the matrix $B$ is used to determine which
row are selected at each iteration step1 of Algorithm 3.2 in a
random experiment to construct a Hamiltonian cycle. In each
independently random experiment running Algorithm 3.2, if $B$ is
chosen, it remains unchanged, which promises the results from the
independent experiment to be identical random variables. \\\\
\textbf{Algorithm 3.2}\\
\textbf{inputs}: $A$, an adjacent matrix of graph $G$;\\
$B=(b_{ij})_{n\times n}$ a matrix,
where $b_{ij}$ is chosen from $\{1,2,\cdots,n-i+1\}$, for any $i,j\in n$. \\
\textbf{outputs}: $X_A$, the estimator of the number of Hamiltonian cycles in $G$;\\
\textbf{step0}: Let $p_i=0$, $i=1,\cdots,n$ and $k=1$.\\
 \textbf{step1}: For $i=1$ to $n$

Set $g_{i}=b_{ik}$ ;

If $|A(g_i,:)| = 1$;

\ \ \ \ \ \ Set $p_n=a_{g_i1}$, goto step2;

Else $W=\{j\neq g_i:a_{g_ij}=1\}$

\ \ \ \ \ \ If $W=\emptyset$; Set $X_A=0$; Stop;

\ \ \ \ \ \ Else

\ \ \ \ \ \ \ \ \ \ Choose $J$ from $W$ uniformly at random;

\ \ \ \ \ \ \ \ \ \ Let $p_{i} =|W|$ and $k=J$;

\ \ \ \ \ \ \ \ \ \ Permutate the column $g_i$ and $J$;

\ \ \ \ \ \ \ \ \ \ Let $A = A(g_i g_i)$;\\
\textbf{step2}: $X_A = p_1 \times\cdots\times p_n$.\\\\
We now show that Algorithm 3.2 presents an unbiased estimator
of the number of Hamiltonian cycles in $G$. Before present the proof we need a technique lemma.\\\\
\textbf{Lemma 3.1} Let $A'(ij)$ denote the matrix obtained from $A$
by removing  row $i$ and column $i$ after permutating column $i$ and
column $j$. Then, if $n\geq 2$
\begin{displaymath}
\operatorname{ham}(A)=\sum\limits_{j\neq
k_1}a_{k_1j}\operatorname{ham}(A'(k_1j)).
 \end{displaymath}\\\\
\textbf{Proof:}
Without loss of generality, we suppose $k_1=1$.\\
The base case $n=2$ is trivial. \\
Suppose the case $n-1$ holds.\\
Now we see the case $n$. Since
\begin{displaymath}
\begin{split}
\operatorname{ham}(A)&=\sum\limits_{\{k_2,k_3,\cdots,k_n\}}a_{1k_2}a_{k_2k_3}\cdots
a_{k_{n-1}k_n}a_{k_n1}\\
&=\sum\limits^{n}_{j=2}\sum\limits_{\{k_3,\cdots,k_n\}}a_{1j}a_{jk_3}\cdots
a_{k_{n-1}k_n}a_{k_n1}\\
&=\sum\limits^{n}_{j=2}a_{1j}\sum\limits_{\{k_3,\cdots,k_n\}}a_{jk_3}\cdots
a_{k_{n-1}k_n}a_{k_n1}
\end{split}
\end{displaymath}
where $\{k_3,\cdots,k_n\}$ is over $\{2,\cdots,n\}/\{j\}$.\\
Hence, we need only to show
\begin{displaymath}
\sum\limits_{\{k_3,\cdots,k_n\}}a_{jk_3}\cdots
a_{k_{n-1}k_n}a_{k_n1}=\operatorname{ham}(A'(1j))
\end{displaymath}
By induction, we know
\begin{displaymath}
\begin{split}
\operatorname{ham}(A'(1j))&=\sum\limits_{\{k'_2,\cdots,k'_n\}}a'_{k'_2k'_3}\cdots
a'_{k'_{n-1}k'_n}a'_{k'_nk'_2}\\
&=\sum\limits_{\{k'_3,\cdots,k'_n\}}a'_{1k'_3}\cdots
a'_{k'_{n-1}k'_n}a'_{k'_n1}
 \end{split}
\end{displaymath}
where $\{k'_3,\cdots,k'_n\}$ is over all the permutations of
$\{2,\cdots,n-1\}$. Recall the definition of $A'(1j)$, which is
obtained by removing  row $1$ and  column $1$ after permutating
column $1$ and column $j$,  then we know $a'_{1k'_3}=a_{j,k'_3+1}$
and $a'_{k'_{3}k'_4}=a'_{k'_3+1,k'_4+1}$, $\cdots$,
$a'_{k'_{n-1}k'_n}=a'_{k'_{n-1}+1,k'_n+1}$,
$a'_{k'_n1}=a'_{k'_n+1,1}$, which completes the proof. \ \ \ $\Box$

By lemma 3.1, it's sufficient to show Algorithm 3.2 presents an
unbiased estimator of the number of Hamiltonian cycles.\\\\
\textbf{Theorem 3.2} Let $X_A$ the output of Algorithm 3.1.
Then $E(X_A)=NH(G)$.\\\\
\textbf{Proof:}
We go on to show $E(X_A)=NH(G)$ by induction on $n$.\\
The base case $n=1$ is trivial.\\
Suppose the case $n-1$ holds, for the case $n$,
\begin{displaymath}
\begin{split}
E(X_A)&=\sum\limits_{j\neq k_1}E(X_A|J=j)P(J=j) \\
      &=\sum\limits_{j\neq k_1}E(p_1X_{A'(k_1j)})/p_1\\
      &=\sum\limits_{j\neq k_1}E(X_{A'(k_1j)}).\\
\end{split}
\end{displaymath}
By induction, $E(X_{A'(k_1j)})=\operatorname{ham}(A'(k_1j))$ and
Lemma 3.1, then
\begin{displaymath}
E(X_A)=\sum\limits_{j\neq
k_1}\operatorname{ham}(A'(k_1j))=\operatorname{ham}(A)=NH(G).
\end{displaymath}
This completes the proof. \ \ \ $\Box$
\section{An Upper Bound of Symmetric Directed Graphs}
We now present the upper bounds for the symmetric directed
graph.\\\\
\textbf{Theorem 4.1} Let $G$ be a symmetric directed graph,
$A=A_{G}=(a_{i,j})_{n \times n}$ be the adjacent matrix of $G$ and
$r_{i}$ denote its sum of row $i$, $i=1,2,\cdots,n$, $n\geq 3$. $N$
denotes the number of Hamiltonian cycles of $G$. Then
\begin{displaymath}
N\leq \frac{1}{2^{n-1}}\prod\limits_{i=1}^{n}r_{i}.
\end{displaymath}
\textbf{Proof:} Let $\mathcal {H}(j)=(m_1, m_2,\cdots, m_{n}, m_1)$
be one of the Hamiltonian cycles of $G$, where $j=1,2,\cdots,N$.
$m_{i}$ $(i=2,3,\cdots,n)$ is a permutation of $\{1,
2,\cdots,n\}/\{m_1\}$. In Algorithm 3.2, choose $b_{11}=m_1$. In
step 1, choose elements by the row order of $m_2,\cdots,m_{n}$ such
that the edges corresponding to the chosen elements constituting
$\mathcal {H}(j)$. Let $S_{i}=\{j:a_{m_ij}=1\}$ and $p^{H}_i$ denote
the value of $i$th iteration of step1 in Algorithm 3.2, where
$i=1,\cdots,n$ and $m_{n+1}=m_1$. Then

$p^{H}_1=r_{m_1}$,

$p^{H}_i=|S{i}/\{m_1,m_2,\cdots,m_i\}|$, $i=2,\cdots,n-1$,

$p^{H}_n=1$.

\noindent Let $X_{\mathcal{H}(j)}$ be the output when the edges
corresponding to the chosen elements form $\mathcal{H}(j)$ , then
$X_{\mathcal{H}(j)}=\prod\limits_{i=1}^{n}p^{H}_i$. Since this is a
symmetric directed graph and $n \geq 3$, there exits a Hamiltonian
cycle $\mathcal {H}^{'}(j)=(m_1, m_n, m_{n-1},\cdots, m_1)$
different from $\mathcal{H}(j)$. Let $p^{H'}_{i}$ be the value of
$i$th iteration of step1 in Algorithm 3.2 and $X_{\mathcal{H}'(j)}$
the output when the edges corresponding to the chosen elements form
$\mathcal {H}^{'}(i)$, where $i=1,\cdots, n$ and $m_{0}=m_{n}$. Then
by Algorithm 3.2,

$p^{H'}_1=r_{m_1}$, $p^{H'}_2=1$,

$p^{H'}_{i}=|S_{i}/\{m_1,m_i,m_{i+1},\cdots,m_{n}\}|$,
$i=3,\cdots,n$,

$X_{\mathcal{H}'(j)}=\prod\limits_{i=1}^{n}p^{H'}_i$.

\noindent Therefore
\begin{displaymath}
X_{\mathcal{H}(j)}X_{\mathcal{H}'(j)}=\prod\limits_{i=1}^{n}p^{H}_ip^{H'}_i.
\end{displaymath}
Considering the symmetry of $A$ and $a_{m_im_i}=0$,
$i=1,2,\cdots,n$, then $p^{H}_i+p^{H'}_i\leq r_{m_i}$,
$i=2,\cdots,n$. Hence
\begin{displaymath}
\begin{split}
X_{\mathcal{H}(j)}X_{\mathcal{H}'(j)}&=\prod\limits_{i=1}^{n}p^{H}_ip^{H'}_i\\
                                     &\leq
                                     r_{m_1}^2\prod\limits_{i=2}^{n}p^{H}_i(r_{m_i}-p^{H}_i)\\
                                     &\leq
                                     \frac{1}{4^{n-1}}\prod\limits_{i=1}^{n}r_{m_i}^2\\                                   \\
                                    &=\frac{1}{4^{n-1}}\prod\limits_{i=1}^{n}r_{i}^2.
\end{split}
\end{displaymath}
Let $X_A$ be the output of Algorithm 3.2. Then by Theorem 3.1 and
Theorem 3.2
$P(X_A=X_{\mathcal{H}(j)})=\frac{1}{X_{\mathcal{H}(j)}}$. From
Algorithm 3.2, we know the output may be zero with certain
probability, hence
$\sum\limits_{j=1}^N\frac{1}{X_{\mathcal{H}(j)}}\leq 1$ . Set
$N=NH(G)$, we have
\begin{displaymath}
\begin{split}
N&\leq \frac{N}{\sum\limits_{j=1}^N\frac{1}{X_{\mathcal{H}(j)}}}\\
     &\leq \sqrt[N]{\prod\limits_{j=1}^{N}X_{\mathcal{H}(j)}}\\
     &=
     \sqrt[2N]{\prod\limits_{j=1}^{N}X_{\mathcal{H}(j)}X_{\mathcal{H'}(j)}}\\
     &\leq
     \sqrt[2N]{\prod\limits_{j=1}^{N}\frac{1}{4^{n-1}}\prod\limits_{i=1}^{n}r_{i}^2}
     =\frac{1}{2^{n-1}}\prod\limits_{i=1}^{n}r_{i}.
\end{split}
\end{displaymath}Thus the result follows. $\Box$\\\\
\textbf{Theorem 4.2} Let $A=(a_{i,j})_{n\times n}$ be an adjacent
matrix of a symmetric directed graph and $r_{i}$ be the sum of row
$i$ of $A$. Then
\begin{displaymath}
\frac{1}{2^{n-1}}\prod\limits_{i=1}^{n}r_{i}\leq
\prod\limits_{i=1}^{n}\frac{r_{i}+1}{2}.
\end{displaymath}
\textbf{Proof:} Since
\begin{displaymath}
\begin{split}
\frac{\prod\limits_{i=1}^{n}\frac{r_{i}+1}{2}}{\frac{1}{2^{n-1}}\prod\limits_{i=1}^{n}r_{i}}
&=\frac{1}{2}(1+\frac{1}{r_i})^{n}\\
&\geq \frac{1}{2}(1+\frac{1}{n})^{n} \geq 1.
\end{split}
\end{displaymath}Thus the result follows. $\Box$\\

Theorem 4.2 shows our upper bound is tighter than Minc's bound (4).
In many cases, the new bound is better than Br\'{e}gman's bound (5).
For example, $$\frac{1}{2^{n-1}}\prod\limits_{i=1}^{n}r_{i}\leq
\prod\limits_{i=1}^{n}(r_i!)^{1/r_i},$$ when $n\geq 100$, $r_i\leq
5$, $i=1,2,\cdots,n$.
\section{Bounds of Undirected Graphs}
The notations or definitions related to the undirected graphs are
only stated in this section. The problem of counting the number of
Hamiltonian cycles in an undirected graph is transformed to that of
counting the number of Hamiltonian cycles in a symmetric directed
graph.  This transformation is very simple but powerful. Let $G$ be
an undirected graph with vertices $\{1,2,\cdots,n\}$, where $n\geq
3$. $G$ is a simple graph. Define a symmetric directed graph $G'$
corresponding to $G$ by replacing each edge $(i,j)$ of $G$ with two
directed edges $(i,j)$ and $(j,i)$. Let $H_G$ and $H_{G'}$ denote
the set of the Hamiltonian cycles in $G$ and $G'$ respectively.
$\mathcal {P}(H_{G'})$ denotes the power set of $H_{G'}$. Recall we
use the notation $v_{1}v_{2}\cdots v_{n}v_{1}$ and
$(v_{1},v_{2},\cdots,v_{n},v_{1})$ to denote Hamiltonian cycles in
undirected  and directed graphs respectively.\\\\
\textbf{Theorem 5.1} Let $\mathcal{H}=m_1m_2\cdots m_{n}m_1$ be a
Hamiltonian cycle in $H_G$. Then there are at least two Hamiltonian
cycles $(m_1, m_2,\cdots, m_{n}, m_1)$ and $(m_1, m_n,
m_{n-1},\cdots, m_1)$ in $H_{G'}$. Define a map $\varphi$ from $H_G$
to $\mathcal {P}(H_{G'})$ as follows:
\begin{displaymath}
\varphi(\mathcal{H})= \{(m_1, m_2,\cdots, m_{n}, m_1), (m_1, m_n,
m_{n-1},\cdots, m_1)\}.
\end{displaymath}
Let $Im\varphi$ denote the image of the map $\varphi$ and
$\mathcal{H}'=m'_1m'_2\cdots m'_{n}m'_1$ be a different Hamiltonian
cycles from $\mathcal{H}$. Then
\begin{displaymath}
\varphi(\mathcal{H}) \cap \varphi(\mathcal{H}')=\emptyset \qquad and
\qquad \cup Im\varphi=H_{G'}.
\end{displaymath}
\textbf{Proof:} Due to the symmetry of the graph and $n\geq 3$, if
there is a Hamiltonian cycle $(m_1, m_2,\cdots, m_{n}, m_1)$ in
$H_{G'}$, there must be a different Hamiltonian cycle $(m_1, m_n,
m_{n-1},\cdots, m_1)$ in $H_{G'}$.  These two Hamiltonian cycles
obviously has a pre-imagine corresponding to  the Hamiltonian cycle
$m_1m_2\cdots m_{n}m_1$ in $H_{G}$. Note $(m_1, m_2,\cdots, m_{n},
m_1)$ is in $\varphi(m_1m_2\cdots m_{n}m_1)$. Hence, $\cup
Im\varphi\supseteq H_{G'}.$ Obviously, $\cup Im\varphi\subseteq
H_{G'}$. Therefore
$$\cup Im\varphi=H_{G'}.$$ Suppose there are two different Hamiltonian
cycles $\mathcal{H}=m_1m_2\cdots m_{n}m_1$ and
$\mathcal{H}'=m'_1m'_2\cdots m'_{n}m'_1$ in $H_{G}$. We know that
they are different iff there exits a vertex $\{m_i\}$=$\{m'_j\}$
such that at least one of the two neighbor vertices of $\{m_i\}$ is
not in the set of two neighbor vertices of $\{m'_j\}$. Hence $(m_1,
m_2,\cdots, m_{n}, m_1)$ is different from $(m'_1, m'_2,\cdots,
m'_{n}, m'_1)$ and $(m'_1, m'_n, m'_{n-1}\cdots, m'_{2}, m'_1)$, we
know $(m_1, m_2,\cdots, m_{n}, m_1)$ is not in the set
$\varphi(\mathcal{H}')$. Similarly, $(m_1, m_n, m_{n-1},\cdots,
m_1)$ is not in $\varphi(\mathcal{H}')$. Then $\varphi(\mathcal{H})$
$\cap$
$\varphi(\mathcal{H}')$ $=\emptyset$. \ \ $\Box$\\\\
\textbf{Corollary 5.1} Let $NH(G)$ and $NH(G')$ denote the the
number of Hamiltonian cycles in undirected graph $G$ and its
corresponding symmetric directed graph $G'$ respectively, then
\begin{displaymath}
NH(G)=\frac{1}{2}NH(G')
\end{displaymath}
\textbf{Proof:} This result is a straightforward deduction of the Theorem 5.1.\ \ $\Box$\\\\
\textbf{Corollary 5.2} Let $G$ be an undirected graph with vertices
$\{1,2,\cdots,n\}$, $n\geq 3$, $NH(G)$ be the number of Hamiltonian
cycles in the graph  $G$. $d_i$ denotes the degree of the vertex
$\{i\}$. Then
\begin{displaymath}
NH(G)\leq \frac{1}{2^{n+1}}\prod\limits_{i=1}^{n}(d_{i}+1),
\end{displaymath}
\begin{displaymath}
NH(G)\leq \frac{1}{2^{n}}\prod\limits_{i=1}^{n}d_{i}
\end{displaymath}
and
\begin{displaymath}
NH(G)\leq \frac{1}{2}\prod\limits_{i=1}^{n}(d_i!)^{1/d_i} .
\end{displaymath}
\textbf{Proof:} By the results of Theorem 2.1, 2.2, 4.1 and
Corollary 5.1,
this corollary follows.\ \ $\Box$\\

\section{Concluding Remarks }

A novel upper bound of the number of Hamiltonian cycles on a
symmetric directed graph is presented first in this paper, which is
tighter than the famous Minc's bound and better than the bound by
Br\'{e}gmman in many cases.

A transformation from the problem of Hamiltonian cycles of an
undirected graph to that of the symmetric directed graph is
constructed. Using this transformation and the bounds for directed
graphs, upper bounds for the number of Hamiltonian cycles in
undirected graph are obtained. The significance  of this
transformation also lies in the fact that the algorithms for
counting the number of Hamiltonian cycles in a directed graph can be
directly applied to count the number of Hamiltonian cycles in an
undirected graph.



\end{document}